\documentstyle[11pt]{article}
\begin{document}
\begin{center}
\begin{Large}
  {\bf Panspermia revisited} \\
\vspace{0.1cm}
  John Gribbin \\
  Astronomy Centre, University of Sussex, Falmer, Brighton BN1 9QJ \\
  J.R.Gribbin@Sussex.ac.uk \\
\end{Large}
\end{center}
\vspace{0.2cm}

     The discovery of evidence for life on Earth more than 3850 million years
  ago (1) naturally encourages a revival of speculation about the possibility
  that life did not originate on Earth, but was carried to the planet in the
  form of microorganisms such as bacteria, either by natural processes or
  deliberate seeding of the Galaxy by intelligent beings.  This idea, known
  as panspermia, has a long history (2, 3), but it is curious that in recent
  decades astronomers have tended to dismiss the possibility of panspermia on
  the grounds that microorganisms could not survive the damage caused by
  ultraviolet radiation and cosmic rays on their journey out of a planetary
  system like the Solar System (4) while some biologists (5) have argued that
  it is impossible for life to have emerged from simple molecules in the
  limited time available (now seen to be substantially less than 1000 million
  years) since the Earth formed.  This has led Crick, in particular, to argue
  that the seeds of life were indeed carried to Earth (and presumably other
  planets) protected inside automated spaceprobes, a process he calls
  directed panspermia (6).

        Recently, however, Wesson and his colleagues (7,8,9,10) have pointed
  out a way in which biological material could escape from a planet like the
  Earth orbiting a star like the Sun by natural processes, and survive with
  its DNA more or less intact.  The problem is that although microorganisms
  could escape from the Earth today, their biological molecules would quickly
  be destroyed by radiation in the near-Earth environment.  Bacteria shielded
  inside fine grains of material such as carbon could survive in the
  interplanetary environment near Earth, but would then be too heavy for the
  radiation pressure of the Sun today to eject them from the Solar System.
  The solution is to argue that suitably shielded microorganisms can be
  ejected from a planetary system like ours when the star is in its red giant
  phase.  This makes it possible for natural mechanisms to seed the Galaxy
  with viable life forms -- and even if the biological material is damaged on
  its journey, as these authors point out, even the arrival of fragments of
  DNA and RNA on Earth some 4000 million years ago would have given a kick
  start to the processes by which life originated here.

        The remaining puzzle about this process is how the grains of
  life-bearing dust get down to Earth.  In their eagerness to suggest how
  microorganisms could have escaped from a planetary system, few of the
  proponents of natural panspermia seem to have worried unduly about how the
  life-bearing grains get back down to a planetary surface.  But the work of
  Wesson and his colleagues naturally leads one to surmise that the immediate
  fate of the microorganisms ejected from a planetary system during the red
  giant phase will be to mingle with the other material ejected from the
  star, forming part of the material of interstellar space and becoming part
  of an interstellar molecular cloud.  When a new planetary system forms from
  such a cloud, it is likely that the accretion processes in he circumstellar
  disc produce very large numbers of cometary bodies, which preserve intact
  the material of the cloud.  Although the processes of accretion of a planet
  like the Earth generate heat which would destroy any microorganisms present
  (and which may well have driven off all the primordial volatiles), it is
  likely that as the planet cools it will be bombarded by comets containing
  large amounts of primordial material (and water) down to the surface (for a
  review, see 11).  If this material includes dormant bacteria, or even
  fragments of DNA, life will be able to get a grip on the planet as soon as
  its surface cools, as seems to have happened on Earth.
 
      The possibility that comets may have brought the seeds of life to Earth
 in this way has been discussed by, for example, McKay (12); but those 
earlier suggestions required that the orghanic mateerial was ejected from
 Earth-like planets inside rocky debris as a result if meteoritic impacts.
 It is difficult to see how material in this form could have become a general
 feature of the interstellar medium, or, indeed, how it would get in to
 comets.  What I propose here, in the light of the work of Wesson and his
 colleagues, is that organic material is not only a natural and widely
 dispersed component of the interstellar medium, but will inevitably be
 incorporated into the material from whoch new planets form.

        The immediate difficulty faced by this hypothesis is explaining why
  life did not get a grip on Venus or Mars, as well -- but that is a
  difficulty shared by all variations on the panspermia theme.  Unlike those
  other variations on the theme, though, this one is testable.  It would be
  feasible to obtain material from a long-period comet, which has never
  previously entered the inner Solar System, and analyse this material for
  traces of DNA.  If the hypothesis is correct, there should be biological
  material very similar to that of life on Earth in these comets.

\vspace{0.2cm}
\noindent {\bf Bibliography} \\

\noindent   (1) Holland, H. D., 1997, Science, 275, 38. \\
   (2) Arrhenius, S., 1908, Worlds in the Making, Harper \& Row, New York.\\
   (3) Shklovskii, I. S. and Sagan, C., 1966, Intelligent Life in the
  Universe, Holden-Day, San Francisco.\\
   (4) Chyba, C. and Sagan, C., 1988, Nature, 355, 125.\\
   (5) Crick, F. H. C. and Orgel, L. E., 1973, Icarus, 19, 341.\\
   (6) Crick, F. H. C., 1982, Life Itself, Macdonald London. \\
   (7) Wesson, P. S., Secker, J., and Lepock, J. R., 1997. Proceedings of the
  5th International Conference on Bioastronomy, IAU Colloquium No. 161, p539,
  Editrice Compositori, Bologna.\\
   (8) Secker, J., Wesson, P. S., and Lepock, J., 1996, Journal of the Royal
  Astronomical Society of Canada, 90, 17.\\
   (9) Secker, J., Lepock, J., and Wesson, P., 1994, Astrophysics and Space
  Science, 219, 1.\\
   (10) Wesson, P. S., 1990, Quarterly Journal of the Royal Astronomical
  Society, 31, 161.\\
   (11) Gribbin, J., in press, Stardust, Viking, London.\\
  (12) McKay, C., 1996, Mercury, 25(6), 15.\\

\end{document}